\documentstyle[epsfig]{article}
\begin{document}

\title{Crystal Barrel Results on Two-Body Decays of the 
Scalar Glueball}
\author{Stefan\,Spanier, on behalf of the Crystal Barrel 
Collaboration\footnote{Email: stefan.spanier@cern.ch}\\
University of Z\"urich, Switzerland; CERN-PPE \vspace{2mm} \\
{\small talk presented at EPS-HEP'97 conference in Jerusalem,
Israel; 19. - 26. August 1997}
}

\maketitle

\begin{abstract}
The Crystal Barrel Collaboration observes scalar meson resonances 
in $\bar p p$ annihilation. Based on the measurements and partial
wave analyses these are candidates for the $^3$P$_0$
groundstate nonet. The supernummerary $f_0(1500)$ resonance
is identified as a scalar groundstate glueball. 
Important information for its characterization comes from the 
decay pattern into pseudoscalar and scalar mesons. 
Data on kaonic decays in the mass region up to 1700~MeV are 
now avaible at Crystal Barrel. New analysis results are presented.
\end{abstract}
\section{The Crystal Barrel Detector}
The Crystal Barrel detector~\cite{detector} was located at the 
Low Energy Antiproton Ring (LEAR) at CERN. Antiprotons with a
momentum of 200~MeV/c were stopped in a liquid hydrogen
target placed in the center of the apparatus.
Here the proton-antiproton annihilation occured.
Measurement of charged tracks was performed with
two cylindrical proportional wire chambers,
which could be replaced by a silicon micro strip detector,
and a jet drift chamber with 23 sensitive wire layers. 
A barrel-shaped calorimeter consisting of 
1380 CsI(Tl) crystals with photodiode readout and coverig 94\% of 
the solid angle 4$\pi$ detected photons from the decay
of neutral mesons like $\pi^0$ and $\eta$ with a
precision in the energy of 2.5\%/$^4\sqrt{E}$.
The assembly was embedded in a solenoid providing a homogeneous
magnetic field of 1.5~T parallel to the incident antiproton beam.
\section{Three pseudoscalar final states}
Suitable channels to explore the scalar mesons are the
three neutral pseudoscalar meson final states.
The reaction proceeds in two steps:
the scalar resonance is produced together with a recoil particle
and in the second step decays into two pseudoscalar mesons.
Scalar mesons
decaying into two pseudoscalar mesons are dominantely produced
from $^1$S$_0$  initial state of the $\bar{p}p$ -system in
liquid hydrogen. Centrifugal barriers hindering the reaction
are absent. A large sample of events triggered
to include only neutral particles which decay into photons
was accumulated. An amount of 712,000 $\pi^0\pi^0\pi^0$,
198,000 $\pi^0\eta\eta$  and 977 $\pi^0\eta\eta^\prime$ 
events could be reconstructed. 
To extract the resonance content in these 
annihilation channels a partial wave analysis was
performed using the K-matrix formalism and
describing the final states simultaneously~\cite{coupled}.
In this analysis model the decay from a 
proton-antiproton initial state into the three final state
particles proceeds successively via intermediate two-body 
states with a certain spin parity $J^{PC}$.
If the intermediate states are resonant they are parametrized
by a mass pole and couplings to the two-body channels. 
A common feature of the all neutral final states mentioned is 
the isoscalar scalar resonance $f_0(1500)$. It could be described
with a common mass and couplings to $\pi\pi$, $\eta\eta$
and $\eta\eta^\prime$.

The Crystal Barrel collaboration observes the $f_0(1500)$
also in the 5$\pi^0$ final state in $\bar{p}p$ annihilation at 
rest~\cite{resag}.
The selection of this 10 photon final state made use of
the all-neutral data sample.
The 4$\pi^0$ invariant mass shows after subtraction of the
$\eta \rightarrow 3\pi^0$ events a peak at a mass of 1450~MeV.
This structure strongly deviates from phase space distribution.
The partial wave analysis explains it as $f_0(1500)$ decaying to
$\sigma\sigma$ ($\sigma$ is a name for the low energy part of the
$\pi\pi$ S-wave) with $\sigma \rightarrow \pi^0\pi^0$ and either produced
together with the 3$\pi^0$ resonance $\pi(1300)$ or produced together
with $f_0(1370)$ and the low mass tail of a resonance above 1700~MeV
in the same S-wave. The $\sigma$ is interpreted as a glue-rich object.

The $f_0(1500)$ is observed in the decay into $K_L K_L$
by studying the final state $K_L K_L \pi^0 $ of annihilation 
at rest~\cite{klkl}.
The $\pi^0$ is fully reconstructed, one $K_L$ is missing and
one $K_L$ undergoes a hadronic interaction in the CsI(TL) crystals
with an average probability of 54\%.
This is sufficient information to reconstruct the kinematics
of an event and to perform a partial wave analysis on the Dalitz plot.
The plot is shown in fig.~1 a).
The resonance features in the $K\pi$ system are
the $K^*(892)$ and the $K^*(1430)$.
In the $\bar{K}K$ subsystem isospin I=0 and I=1 is possible.
Therefore $f_2(1270)$ and $a_2(1320)$ are seen together.
The $f_2^\prime(1525)$ adds to the $I=0$ $\bar{K}K$ D-wave.
A strong contribution of the $\bar{K}K$ S-wave is found. 
At least two poles in a $1\times 1$ K-matrix
are needed in the S-wave to arrive at a satisfactory description 
of the data. These belong to the I=0 resonances $f_0(1370)$ and 
$f_0(1500)$. The result stays ambiguous since one also expects 
the presence of the $a_0(1450)$ resonance which was observed in 
its $\pi^0\eta$ decay~\cite{ppe}.
Any contribution of the $a_0(1450)$ between 0\% and
15\% does only affect the $\bar{K}K$ S-wave but not the quality of 
the fit.

To resolve the isospin ambiguity the final state $K_L K^\pm\pi^\mp$
of $\bar{p}p$ annihilation at rest was selected~\cite{klkp}.
In the reaction $\bar{p}p \rightarrow K_L K^\pm \pi^\mp$ only the
$I=1$ $\bar{K}K$ resonances are produced. By applying isospin
symmetry one can calculate their contributions to the $K_L K_L \pi^0$
channel.
The $K_L K^\pm \pi^\mp$ final state was reconstructed by requiring
a missing $K_L$ and two charged particles, which
are identified by $dE/dx$~\cite{klkp}. 
An amount of 11,373 events went into
the Dalitz plot displayed in fig.~1 b).
The branching ratio for the proton antiproton annihilation at rest 
into this final state is found compatible with earlier
bubble chamber determinations on less statistics. The average is:
$BR(\bar{p}p \rightarrow K_LK^\pm\pi^\mp)$ = 
$2.74\pm 0.10)\cdot 10^{-3}$~\cite{klkp}.
The total fraction of background from other annihilation channels
is below 2\%. The partial wave analysis revealed necessity
for the introduction of the $I=1$ resonance $a_0(1450)$.
Mass and width were determined as $m=(1480\pm 30)$~MeV
and $\Gamma = (265\pm15)$~MeV, respectively. 
The comparison with the annihilation channel $\pi^0\pi^0\eta$ 
yields the relative ratio 
$B(a_0\rightarrow \bar{K}K)/B(a_0\rightarrow \pi\eta)$ = $0.88\pm0.23$
which agrees well with the prediction from SU(3) flavour symmetry
assuming that this object is member of the scalar nonet.
Having determined its contribution the $f_0(1500)$ decay
to $\bar{K}K$ can be fixed. The relation is shown in 
fig.~\ref{mhres}. The branching ratio for $f_0(1500)$ from
this combined analysis is: \\
$B(\bar{p}p \rightarrow \pi f_0(1500); f_0(1500)\rightarrow \bar{K}K)$ =
$(4.52\pm 0.36)\cdot 10^{-4}$.
\begin{figure}
\begin{tabular}{cc}
\mbox{\epsfig{file=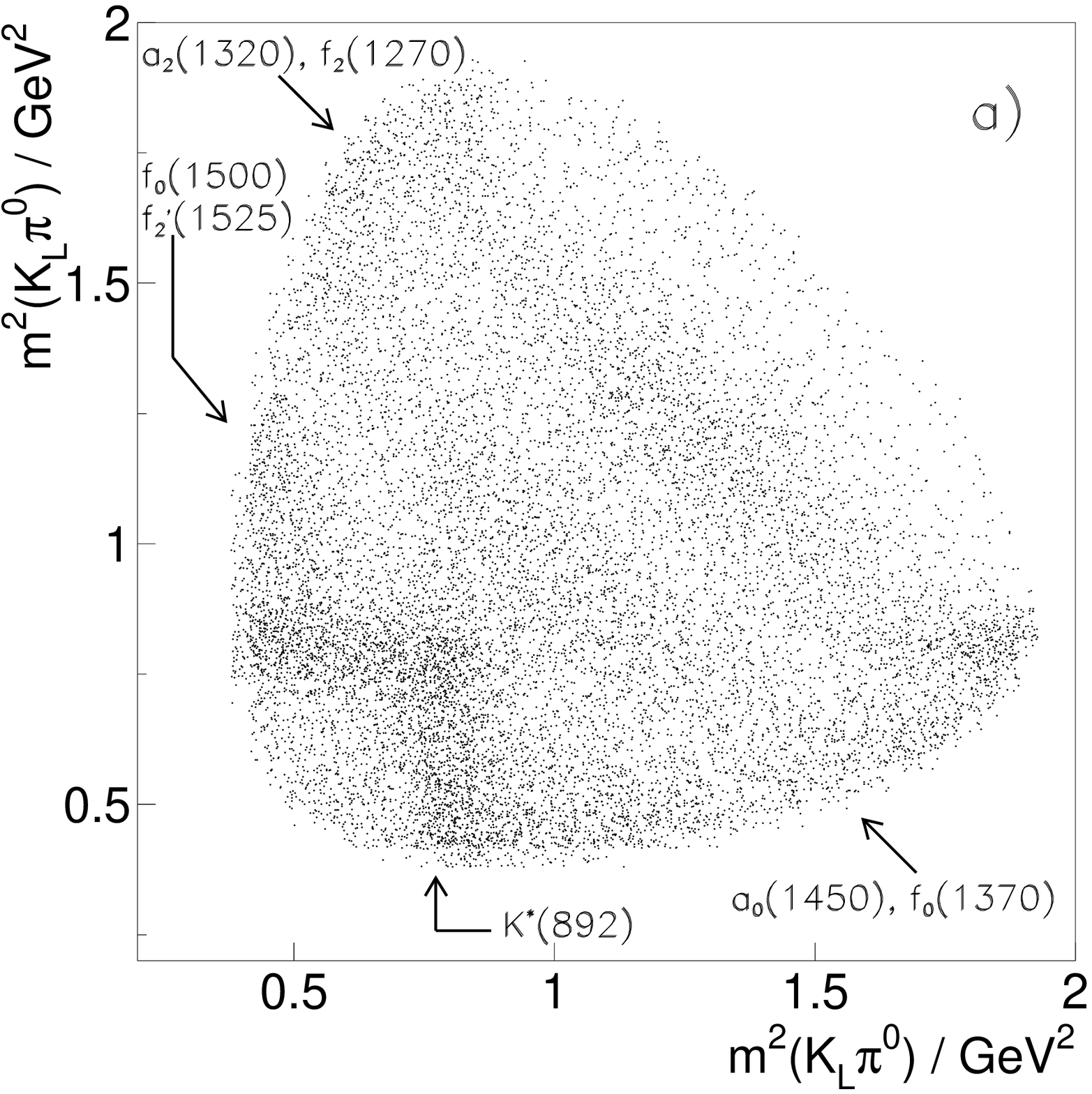,height=6.5cm}} 
\mbox{\epsfig{file=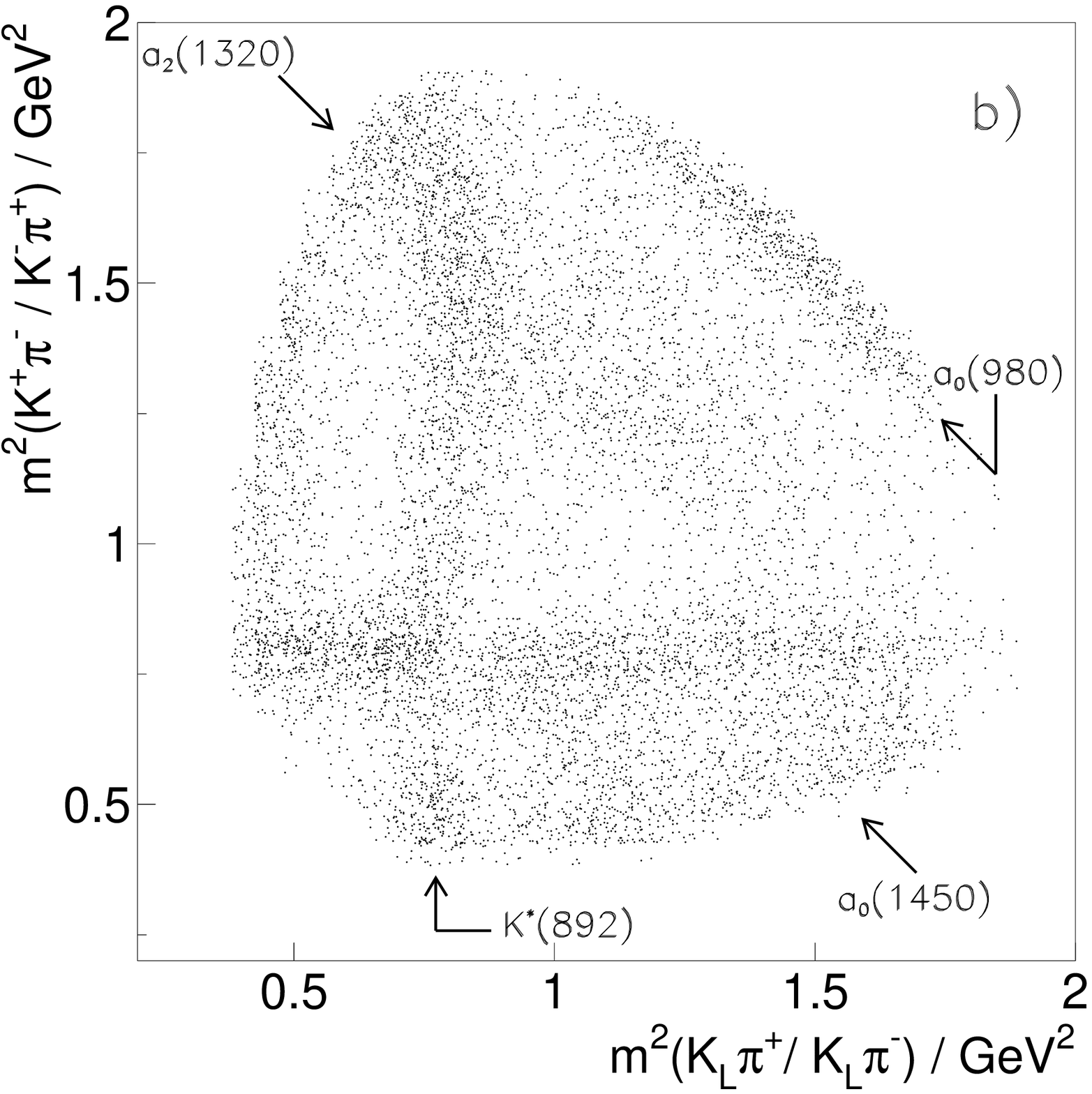,height=6.5cm}} 
\end{tabular}
\label{dalis}
\caption{The Dalitz plots for the annihilation reactions
a) $\bar{p}p \rightarrow K_L K_L \pi^0$ and 
b) $\bar{p}p \rightarrow K_L K^\pm \pi^mp$ in liquid hydrogen.}
\end{figure}
\begin{figure}
\mbox{\epsfig{file=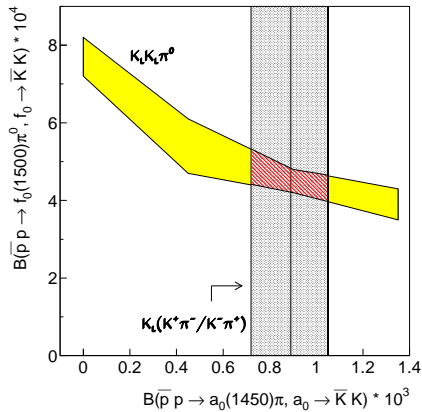,height=6.0cm}}
\label{mhres}
\caption{Determination of the branching ratio of $f_0(1500)$ in the
reaction $\bar{p}p \rightarrow K_L K_L \pi^0$~\cite{klkl}
(light hatched area), where it is correlated with the $a_0(1450)$ 
production. The $a_0(1450)$ contribution is fixed in the analysis
of the reaction $\bar{p}p \rightarrow K_L K^\pm \pi^mp$~\cite{klkp}.}
\end{figure}
\section{Summary}
The phase space corrected couplings of $f_0(1500)$ are: \\
$\pi\pi : \eta\eta : \eta\eta^\prime : \bar{K}K = 
1 : 0.25\pm0.11 : 0.35\pm0.15 : 0.24\pm 0.09$.
Due to these couplings and the strong $\sigma\sigma$ decay 
the $f_0(1500)$ appears as $\omega$-like member of the 
scalar meson nonet. Other candidates for a nonet in this 
mass range are the $I=0$ $f_0(1370)$,
$I=\frac{1}{2}$ $K^*(1430)$  and $I=1$ $a_0(1450)$. 
The mass of $m = 1505\pm 9$~MeV would fit
but the $I=0$ nonet position is also occupied by the $f_0(1370)$.
The width $\Gamma=111\pm12$~MeV of $f_0(1500)$
is relatively small in comparison to the other scalar mesons 
having $\Gamma$ greater than 250 MeV. 
Hence, it appears supernummerary.
The $f_0(1500)$ matches the mass range of 
lattice gauge calculations for the scalar
groundstate glueball~\cite{lqcd}.
The strong coupling to $\eta\eta$, $\eta\eta^\prime$ 
and $\sigma\sigma$ can be understood
in terms of the decolorization mechanism: The constituent gluons
couple to the glue-content of the decay mesons and are color-neutralized
afterwards by the exchange of gluons~\cite{gersh}.
The naive expectation of the flavour democratic decay of a
glueball can be explained by the mixing with nearby
$\bar{q}q$ meson states~\cite{closeams}\cite{farrar}.
The strength of the mixing depends on the mass of the $\phi$-like state
which therefore is predicted at higher mass. 
To explore the mass region above 1700~MeV  
the Crystal Barrel Collaboration is analyzing annihilation 
channels at higher momenta of the incoming antiprotons.
%
%

%
\end{document}